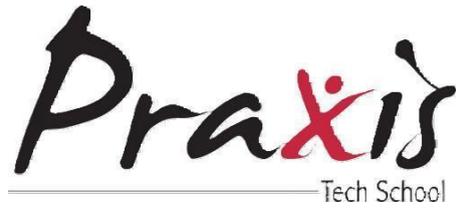

# Evaluating Adversarial Robustness: A Comparison Of FGSM, Carlini-Wagner Attacks, And The Role of Distillation As Defense Mechanism

A Capstone Project report submitted in partial fulfillment of the requirements for the Post Graduate Program in Data Science at Praxis Tech School, Kolkata, India

By

**Trilokesh Ranjan Sarkar (Roll No: A23047)**

**Nilanjan Das (Roll No: A23028)**

**Pralay Sankar Maitra (Roll No: A23029)**

**Bijoy Some (Roll No: A23013)**

**Ritwik Saha (Roll No: BM22038)**

**Orijita Adhikary (Roll No: BM22053)**

**Bishal Bose (Roll No: A23015)**

of Data Science 2023 Fall Batch

Under the supervision of

**Prof. Jaydip Sen**
**Praxis Tech School, Kolkata, India**

# 1. Introduction

As artificial intelligence progresses, concerns over adversarial attacks - deliberate manipulations of machine learning models through altered data - are mounting. This article delves into the investigation of adversarial attacks specifically targeting Deep Neural Networks (DNNs) used for image classification. We focus on understanding the impact of two prominent attack approaches: the Fast Gradient Sign Method (FGSM) and the Carlini-Wagner (CW) approach, on three pre-trained image classifiers. Adversarial attacks pose a significant threat to the reliability and security of machine learning systems, particularly in critical applications like image classification. These attacks involve subtly manipulating input data to cause misclassification by the model, leading to potentially harmful consequences [21]. The FGSM attack is a widely studied method that perturbs input images by leveraging gradients of the loss function to generate adversarial examples efficiently [1].

However, the CW approach represents a more sophisticated class of attacks, developed by Carlini and Wagner, which surpasses earlier techniques in terms of success rates with minimal perturbations [6]. The CW attack formulates the adversarial example generation as an optimization problem, utilizing powerful gradient-based algorithms such as L0, L2, and L∞ norms, to find the smallest perturbation necessary for misclassification [22]. Specifically, the L0 norm counts the number of non-zero elements in the perturbation, the L2 norm measures the Euclidean distance between the original and perturbed input, and the L∞ norm calculates the maximum absolute difference between corresponding elements of the original and perturbed input [23]. This advanced nature of the CW attack presents additional challenges for defense tactics, as it can bypass traditional defense mechanisms designed to mitigate attacks like FGSM [24].

In the face of escalating adversarial threats, the research community has been actively exploring various defense mechanisms to bolster the robustness of machine learning models. Defensive distillation, a technique that trains a model on softened probabilities from a pre-trained model, has emerged as one such strategy. This approach aims to enhance the model's resilience against adversarial perturbations. However, evaluations indicate that even with this modified defensive distillation method, there is a notable decline in performance when contending with sophisticated attacks like the CW attack [11]. While defensive distillation has demonstrated potential in mitigating simpler attacks such as the FGSM, its efficacy is reduced against more complex techniques [1].

The intricacies of the CW attack highlight the necessity for defense mechanisms that can effectively counteract gradient-based optimization strategies. This may entail the development of innovative defense strategies that take into account the unique characteristics of advanced attacks like CW and adaptively modify the model's architecture or training process to improve resilience [6]. In summary, adversarial attacks on Deep Neural Networks (DNNs) for image classification pose a formidable challenge in artificial intelligence. Defense mechanisms like defensive distillation provide a measure of protection but are not infallible and may falter against sophisticated techniques like CW. Future research should concentrate on devising more robust defense strategies that can counter the effects of advanced adversarial attacks, thereby ensuring the reliability and security of machine learning systems in critical applications [8].

# 2. Related Work

Significant progress has been made in several domains since the introduction of deep learning, such as autonomous systems, natural language processing, and picture and audio recognition. The foundation of these advancements has been established by neural networks, especially Deep Neural Networks (DNNs), which are capable of learning intricate patterns and producing remarkably accurate predictions. Nonetheless, these networks' vulnerability to hostile attacks presents a serious threat to their dependability and security.

**Types of Adversarial Attacks**

1. **FGSM Attack**: The FGSM [1] assault was first presented as a straightforward but powerful way to produce adversarial examples. To cause misclassification, the approach uses the gradient of the loss function concerning the input data to perturb the input in a direction that maximizes the loss. Deep neural networks (DNNs) have flaws when they show how subtle changes to input photos could cause misclassification [2]. They emphasized the necessity of resilience to these hostile assaults. Extending FGSM to the Jacobian-based Saliency Map Attack (JSMA) [3] introduced a more focused method for producing adversarial cases. They underlined how crucial it is to assess how resilient machine learning models are to hostile attacks.

2. **Patch Attack:** Patch attacks include applying a carefully designed patch to an image, which can lead to the patched image being incorrectly classified by state-of-the-art classifiers. These attacks pose a threat to real-world systems since they are not limited to being effective in digital space, but they may also be physically executed. In many real-world applications, such as driverless cars and medical diagnostics, image classifiers are essential. Recent studies have brought to light these systems' weaknesses, notably in relation to their vulnerability to hostile attacks. Patch assaults are one type of attack that has gained a lot of attention lately. These attacks involve intentionally placing small patches that might lead to misinterpretation. In this study, we look at relevant work on patch attacks on image classifiers, with an emphasis on the techniques created to produce these kinds of attacks.

   1. **Single-Object Patch Attacks:** This method involves adding a tiny patch to a picture, usually aimed at a single object in the scene. Evolutionary methods could be used to maximize misclassification while minimizing visual distortion in the patch. [4].
   2. **Universal Patch Attacks:** Universal Adversarial Perturbations (UAPs) are designed to develop patches that can mislead a classifier across several photos, as opposed to optimizing patches for specific images. UAPs [5], also showed how well they worked to produce subtle perturbations that led to the misclassification of a variety of images.

3. **CW Attack:** An adversarial strategy noted for its success against models that have defensive measures in place is the CW attack, which is optimization-based. It deliberately creates small-scale perturbations with great care that are intended to produce misclassifications with high confidence. The CW attack was first presented by Carlini and Wagner, who also showed how well it worked to produce adversarial samples for a variety of machine-learning models. They put forth an optimization-based method to identify the smallest perturbations that result in misclassification, and it allows the attack's intensity to be adjusted using various distance measures [6]. By putting forth the robust optimization framework, [7] expanded on the CW approach and improved the transferability of adversarial examples across other models and datasets. To improve the resilience of created adversarial examples, they introduced a novel objective function that promotes the perturbations to lie within a narrow zone surrounding the original input. Adversarial training has been studied as a defence mechanism against adversarial attacks to improve robustness [8]. They trained models using adversarially altered samples. Their research demonstrated how well adversarial training using projected gradient descent may strengthen deep learning models' defenses against a range of attacks, such as the CW attack. By adding a momentum factor to speed up convergence during optimization, improved upon previous attacks like CW and introduced the momentum iterative approach [9]. Compared with conventional iterative methods, their methodology achieved greater success rates and indicated enhanced effectiveness in producing adversarial examples. To give a thorough assessment of model vulnerabilities, an ensemble-based approach for evaluating adversarial robustness was proposed [10]. This approach combined various varied attacks, including CW. By taking into account a variety of attack techniques, their method enables a more dependable assessment of model robustness without the need for extra hyperparameters.

4. **Other Adversarial Attacks:** "DeepFool is a simple and accurate method to fool deep neural networks." To minimize disturbance while misclassifying an image, this method iteratively moves the input image in the direction of the decision boundary [19]. An effective iterative adversarial assault technique for creating adversarial instances is the Projected Gradient Descent (PGD) assault. Subject to a maximum permitted perturbation size, which is usually expressed in terms of the L∞ norm, the goal is to identify the perturbation that maximizes the loss function [8].

**Defensive Strategies**

The tactics used to fight off hostile attacks change along with them. Numerous protective strategies have been developed in the area to increase neural networks' resilience.

**Defensive Distillation:** By using soft labels from a previously trained model to train a new model, a technique known as defensive distillation can strengthen the model's resistance to attacks such as FGSM [31]. The model's output surface is effectively smoothed by this approach, which makes it more difficult for gradient-based attacks to identify successful perturbations. A method called "defensive distillation" was put forth [11] to strengthen neural networks' resistance to hostile attacks. A model is trained using softened labels created by another model that was trained using the same set of data. Its effectiveness has been questioned, though, in light of more complex attacks like the CW attack, which contrasts with its triumph over less complex ones like the FGSM. Defensive distillation was presented [11] to strengthen neural networks' resistance to hostile attacks. They suggested using temperature scaling to train a second model with softened labels produced by the first model, hence lowering prediction confidence. At first, this approach seemed to have the potential to strengthen the model's resistance to attacks such as FGSM. In-depth tests were carried out to assess defensive distillation's resilience to hostile assaults [12]. Defensive distillation proved to be ineffectual against more complex attacks such as the CW attack, but it proved to be resilient against less complex ones like FGSM. The limitations of defensive distillation in countering sophisticated hostile methods were brought to light by this study. This work investigated deep learning models' vulnerabilities in hostile environments in more detail. They discovered that robustness against adaptive adversaries using strategies like the CW attack is not provided by defensive distillation. The necessity for stronger defense systems that can withstand cunning adversary tactics was highlighted by this study. Defensive distillation faces a major challenge from the CW assault, which was first described in Carlini and Wagner's groundbreaking work on evaluating neural network resilience. Their research exposed defensive distillation's shortcomings in actual adversarial situations by showing that, although strong against less sophisticated attacks, it is not a reliable defense against optimization-based attacks such as CW [6].

To sum up, defensive distillation has been demonstrated to be ineffective against more complex attacks like the CW attack, whilst originally showing promise in boosting model resilience against adversarial attacks like FGSM [32]. The significance of creating more all-encompassing defense mechanisms to lessen deep learning models' weaknesses in hostile environments is highlighted by these findings.

**Datasets**

- **MNIST:** MNIST is a commonly used dataset made up of handwritten numbers (0–9) in 28x28 grayscale photos. Because of its simplicity and ease of experimentation, it is a popular choice for benchmarking image classification algorithms. It contains 10,000 testing photos in addition to 60,000 training images [15].

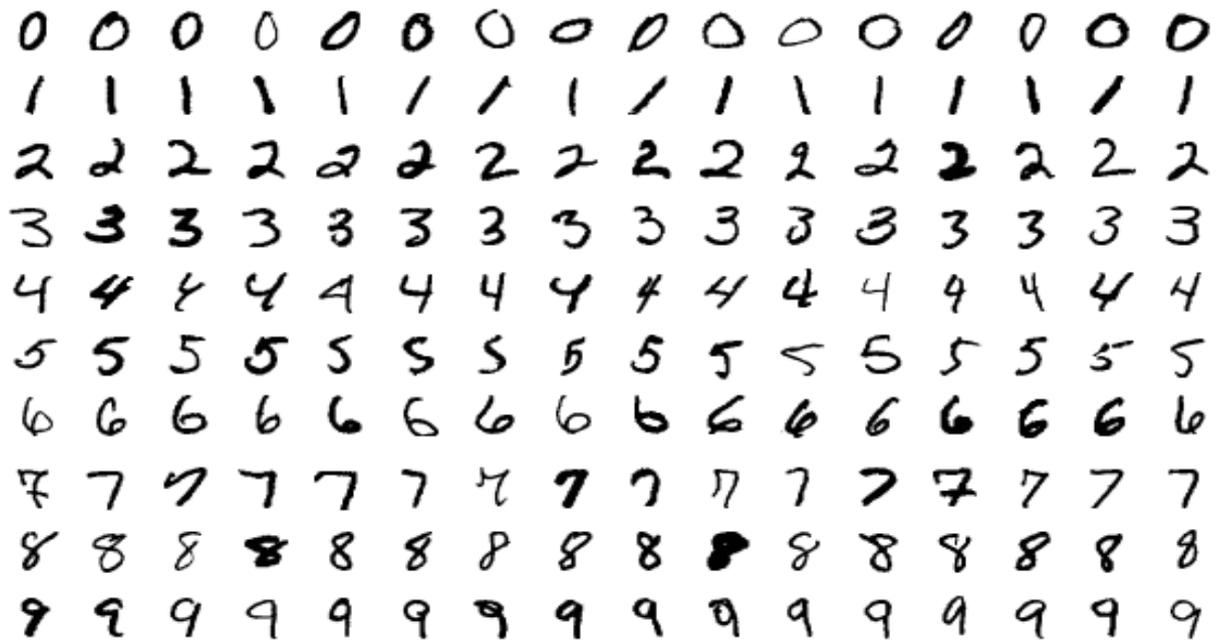

*Fig. 1: MNIST Handwritten Dataset*

- **CIFAR10:** There are 60,000 32x32 color images in CIFAR-10, with 6,000 images in each of the ten classes. It is harder for picture classification jobs since it covers a wider range of things than MNIST, such as automobiles, animals, and common objects [16].

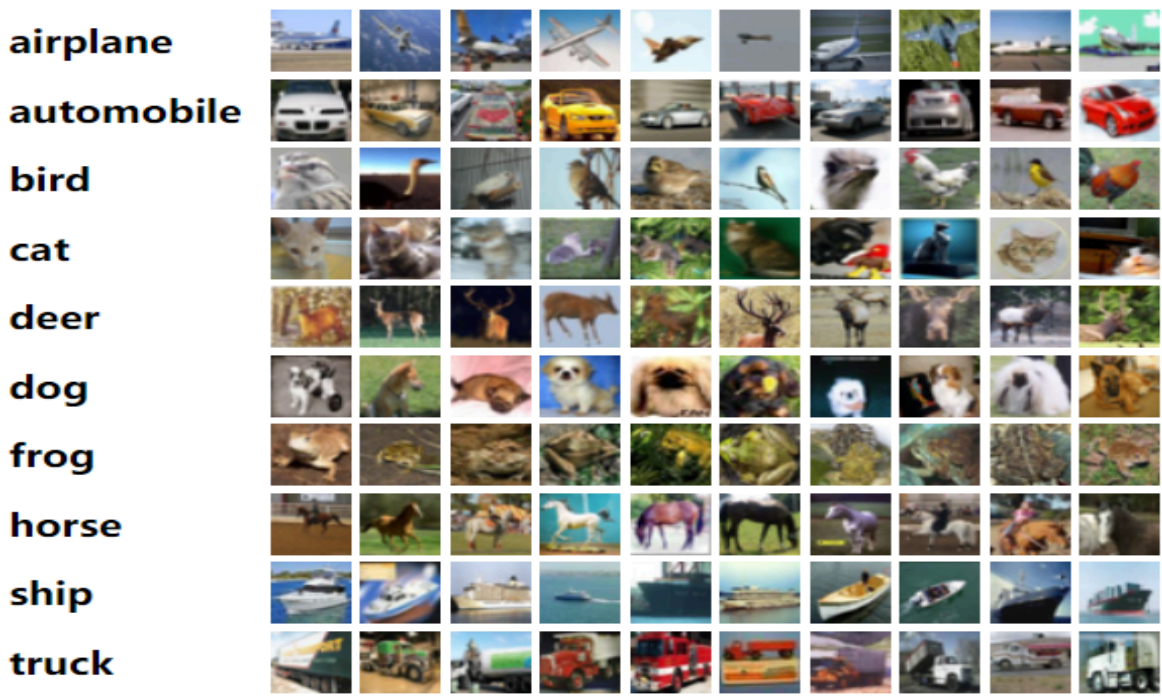

*Fig. 2: CIFAR-10 Dataset*

- **CIFAR100:** CIFAR-100 is an expansion of CIFAR-10 that has 600 photos in each class and 100 classes in total. Every image retains its RGB format and dimensions of 32 by 32 pixels. In comparison to CIFAR-10, CIFAR-100 offers a higher degree of difficulty and a wider range of object classes [16].

- **ImageNet:** One of the biggest and most used datasets for image classification applications is ImageNet. It has more than 20,000 categories and more than 14 million photos. Deep learning models may be trained and evaluated on a huge scale because of the dataset's extensive range of objects and situations [17].

- **Tiny-ImageNet:** With 200 object classes and 500 training photos per class, Tiny ImageNet is a condensed version of the original ImageNet dataset. Every image has a size of 64 by 64 pixels. By acting as a bridge between larger ImageNet and smaller datasets such as CIFAR, tiny ImageNet offers a more difficult standard for image classification applications [18].

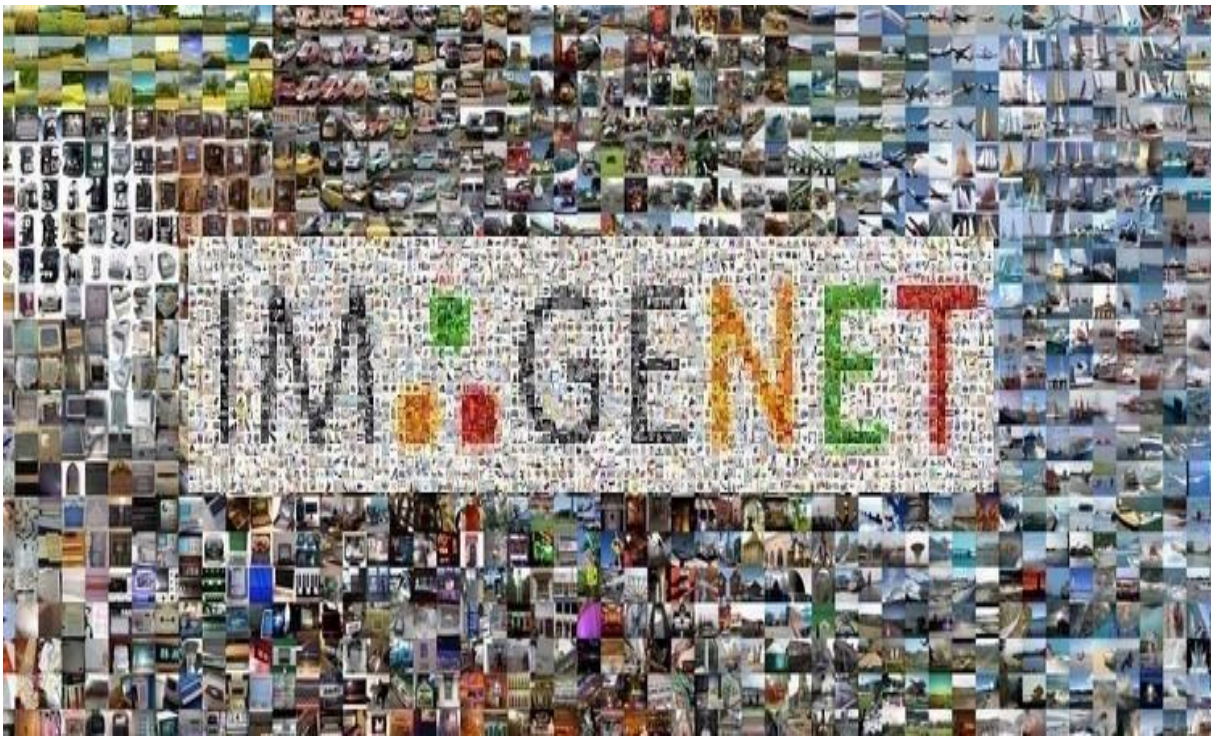

*Fig. 3: ImageNet Dataset*

# 3. Methodology

In addition to examining the feasibility of Defensive Distillation as a potential defense tactic, the study conducted a thorough investigation into the vulnerability of three CNN models, Resnext50_32x4d, DenseNet201, and VGG19, to adversarial attacks. We can outline the research approach used in this study using a step-by-step breakdown.

Initially, PyTorch's torchvision package, which made pre-trained CNN models accessible, was used to rigorously evaluate the models. The Tiny ImageNet dataset was used to carefully evaluate these models to define baselines for essential performance. The study sought to measure the models' intrinsic ability to perform picture classification tasks by computing key classification accuracy metrics, such as Top-1 and Top-5 mistakes. Furthermore, a portion of the photos was carefully chosen for close examination, providing insightful information about the models' decision-making procedures and classification results.

The study next turned its attention to adversarial attacks, utilizing two well-known techniques: the CW attack and the FGSM. The study aimed to investigate the effect of perturbation magnitude on model susceptibility by systematically altering the epsilon values, ranging from 1% to 10%, for both attack approaches. By investigating classification accuracy metrics in detail and documenting classification mistakes at various epsilon values, the study sought to reveal the complex behavior of the models under adversarial pressure.

Additionally, the research investigated Defensive Distillation as a possible countermeasure against hostile assaults. A ResNet101 model was first trained on the CIFAR-10 dataset to capture a plethora of knowledge using a teacher-student structure. Later, this information was condensed into a smaller student model, the Resnext50_32x4d architecture. Through the evaluation of the student model's accuracy before and during FGSM attacks at different epsilon values, the research examined the effectiveness of Defensive Distillation in reducing the negative consequences of adversarial perturbations.

To put it briefly, the study technique used here was multimodal, ranging from a thorough model review to the simulation of adversarial attacks and the evaluation of defense mechanisms that followed. The goal of the study was to obtain important insights about CNN architectural weaknesses and defense strategy efficacy in the complex field of deep learning through these painstakingly constructed assessments.

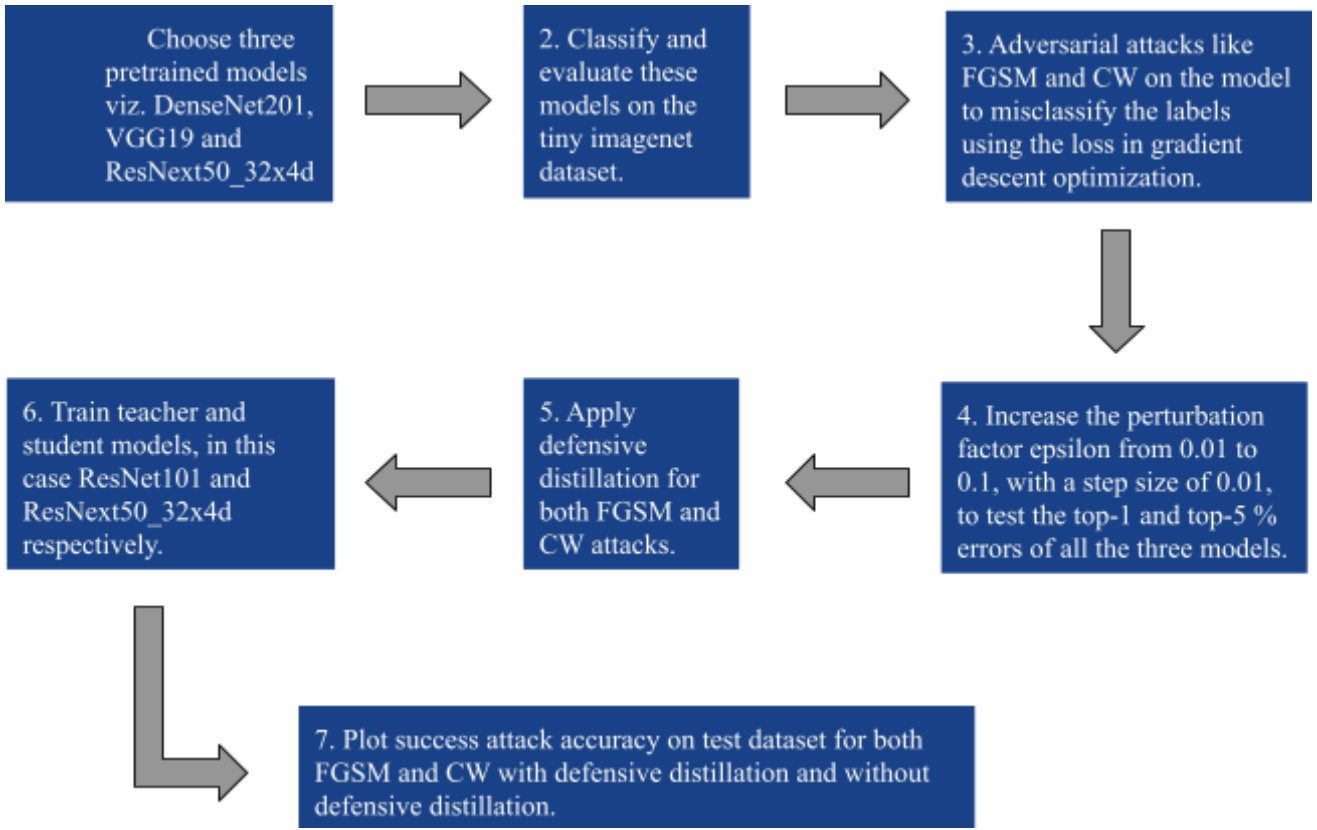

*Fig. 4: Flowchart depicting the methodology section of our research project*

## 4. Background Theories and Model Architecture

**FGSM Attack**

The FGSM [1] assault is a straightforward yet powerful way to produce adversarial cases. It makes use of the loss function's gradients to produce perturbations that maximize loss and cause misclassification.

Given an input image x, a neural network model f with parameters θ, a loss function L, and a small perturbation magnitude ϵ, the adversarial example $x_{adv}$ is computed as:

$$x_{adv} = x + \epsilon \cdot sign(\nabla_x L(f(x; \theta), y_{true}))$$

Where,

- $x_{adv}$ is the adversarial image,
- x is an original image,
- ϵ is a small scalar representing the magnitude of the perturbation,
- $\nabla_x L(f(x; \theta), y_{true})$ is the gradient of the loss function L concerning the input image x,
- $y_{true}$ is the proper label of the input image,
- sign(·) denotes the sign function, which extracts the sign of each element of the gradient vector.

**CW Attack**

The CW attack is an adversarial attack technique that solves an optimization issue to provide undetectable adversarial samples. It seeks to identify the lowest perturbation that causes misclassification while still being undetectable under a given distance measure. Through an iterative optimization process that strikes a compromise between perturbation magnitude and misclassification loss, the C&W approach produces adversarial samples that are highly effective and challenging for neural networks to identify.

The theory behind the CW attack[6] was explained as, letting a perturbation, represented as δ, is introduced for a given image x, aiming to minimize the distance metric D(x; x + δ) when added to x. This perturbation is subject to the condition that the resulting image x + δ is classified as the target class t. The objective is to achieve a subtle modification of x to induce a change in its classification while ensuring the perturbed image remains recognizable and valid. However, solving this problem directly is challenging due to the complex non-linear constraint C(x + δ) = t.

To tackle this challenge, an alternative approach is proposed. In this formulation, an objective function is defined, combining the distance metric D(x; x + δ) with a regularization term f(x + δ), scaled by a positive constant c. This modified formulation aims to simplify the optimization problem. Importantly, the equivalence between the original and alternative formulations suggests the existence of an appropriate constant c ensuring the optimal solution of the alternative problem corresponds to that of the original one.

By substituting the distance metric D with an Lp norm, the problem is reformulated as minimizing ( δ + c * f(x + δ) ), where p indicates the chosen norm.

So in brief after substituting the distance D with an Lp norm, the issue transforms into:

$$minimize\ |\delta|_p + c \cdot f(x + \delta)$$

subject to the constraint:

$$x + \delta \in [0, 1]^n$$

where:

- $|\delta|_p$ denotes the $L_p$ norm of $\delta$,

- $c$ is a positive constant,

- $f(x + \delta)$ represents the regularization term,

- $x + \delta$ belongs to the valid image space $[0, 1]^n$.

**Defensive Distillation**

Defensive Distillation was presented as a defense mechanism to strengthen machine learning models' resilience against adversarial attacks. Instead of using the hard labels directly, it entails training a model using a softened version of the output probabilities produced by a pre-trained model. The field of knowledge distillation, where the goal is to transfer knowledge from a large, complicated model (teacher) to a smaller, simpler model (student), is where the term "distillation" originated.

The two primary steps in the defensive distillation training method are as follows:

● Pre-training: Using conventional supervised learning methods, a sizable, well-trained model (called the teacher model) is first trained on the relevant dataset. This model provides the knowledge base for the defensive distillation procedure since it has been trained to generate precise predictions on the dataset.

- Distillation Training: After that, a smaller model (the distilled model) is trained on the same dataset, but it does so by using the teacher model's softened probabilities as training inputs rather than the actual hard labels. By adding a temperature parameter to the instructor model's softmax output, these softened probabilities are produced, which leads to a more diffuse and smooth probability distribution. Based on these softer probabilities, the distilled model is educated to imitate the instructor model's actions.

The objective during defensive distillation training is to minimize the overall loss, which is a combination of the classification loss and the distillation loss. Mathematically, the training objective can be expressed as:

$$\text{Total Loss} = \text{Classification Loss} + \lambda \times \text{Distillation Loss}$$

Where $\lambda$ is a hyperparameter that controls the importance of the distillation loss relative to the classification loss.

Defensive distillation is justified by the fact that the smoothed probabilities give the student model a more reliable and steady signal to work with, which reduces its susceptibility to minute changes in the input data caused by hostile attacks. The student model becomes more tolerant to small fluctuations that may arise from adversarial perturbations by learning to focus on the most prominent characteristics of the data by training on the softer probabilities.

**ResNext50_32x4d Model**

The convolutional neural network (CNN) model Resnext50_32x4d [25] contains 50 layers and 32 x 4 dimensions. This Model belongs to the ResNeXt family, which is an extension of the ResNet (Residual Network) architecture. The "Next" in ResNeXt [28] refers to the concept of "Next" or "Next Dimension." This is achieved through the introduction of a new module called a "cardinality" module, which incorporates grouped convolutions. The ResNeXt-50 model is constructed by a template with cardinality = 32 and bottleneck width = 4d. The "32x4d" part of the model name refers to the cardinality and base width parameters. Cardinality refers to the number of groups in the grouped convolutions. In ResNeXt50_32x4d, the cardinality is set to 32, meaning that each convolutional layer divides its input into 32 groups and performs separate convolution operations within each group, and "4d" represents the base width of the bottleneck layer. The bottleneck layer has a base width of 4, meaning each group in the bottleneck layer has 4 channels. Since the cardinality is 32, there are 32 groups in the bottleneck layer.

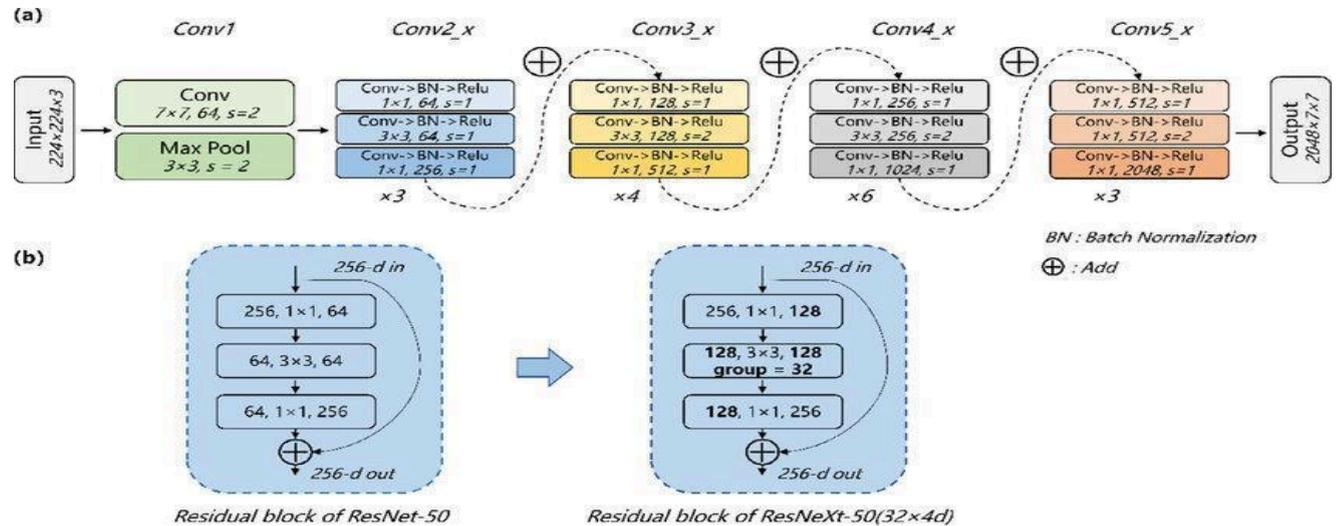

Fig. 5: Resnext50_32x4d network architecture

**DenseNet201 Model**

The DenseNet-201 [24,27] consists of a 201-layer convolutional neural network with 20,242,984 parameters. In DenseNet-201 [30], there are 98 blocks of densely connected layers, including both 1x1 and 3x3 convolutional layers. A globally average pooling layer and a fully connected layer come after these blocks. The network is pre-trained, having been trained on more than a million photos, using the ImageNet database Images of objects such as keyboards, mice, pencils, and other animals will be classified by the network into 1000 distinct categories. Consequently, the network has picked up comprehensive feature representations for a range of picture kinds. The key idea behind DenseNet is to allow each layer to directly access the outputs of all preceding layers, making information flow more efficiently. This approach reduces the number of parameters needed compared to traditional CNNs, which helps save memory and speed up computations. In the DenseNet201 model, the last layer utilizes a SoftMax activation to ascertain the classification class. Even though DenseNet's structure might seem complex, it offers better performance in tasks like image recognition.

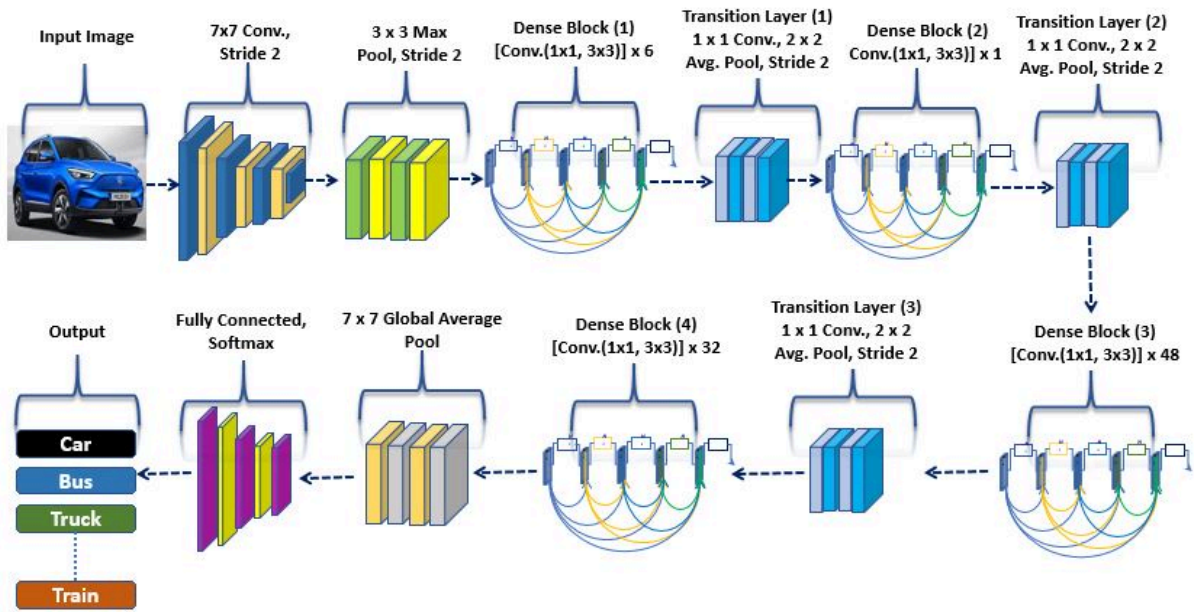

*Fig. 6: DenseNet201 network architecture*

**VGG 19 Model**

The VGG [23] architecture is a well-known model for its ability to classify images effectively. There are five blocks including sixteen convolution layers in this model. Following each block comes the Maxpool layer, which compresses the input image's size by two and doubles the filters of the convolution layer. It uses blocks of convolutional layers, where each block includes 3x3 filters, 1x1 padding, and 2x2 max-pooling. VGG-19[26], a version with 19 layers, stands out with 143 million parameters, setting a standard for CNN performance. These convolution layers extract features from images, enhanced by ReLU activation for recognizing complex patterns. Max-pooling then reduces complexity while keeping important details. Dropout layers help prevent overfitting by randomly turning off some neurons during training. ReLU activation also helps by addressing the vanishing gradient problem, making training more efficient. In this model, the last layer typically consists of a SoftMax activation function. The SoftMax layer takes the output of the preceding fully connected layers and computes the probabilities for each class, enabling the model to classify images effectively based on the highest probability class. Finally, these fully connected layers at the end allow for high-level decision-making and accurate classification based on learned features.

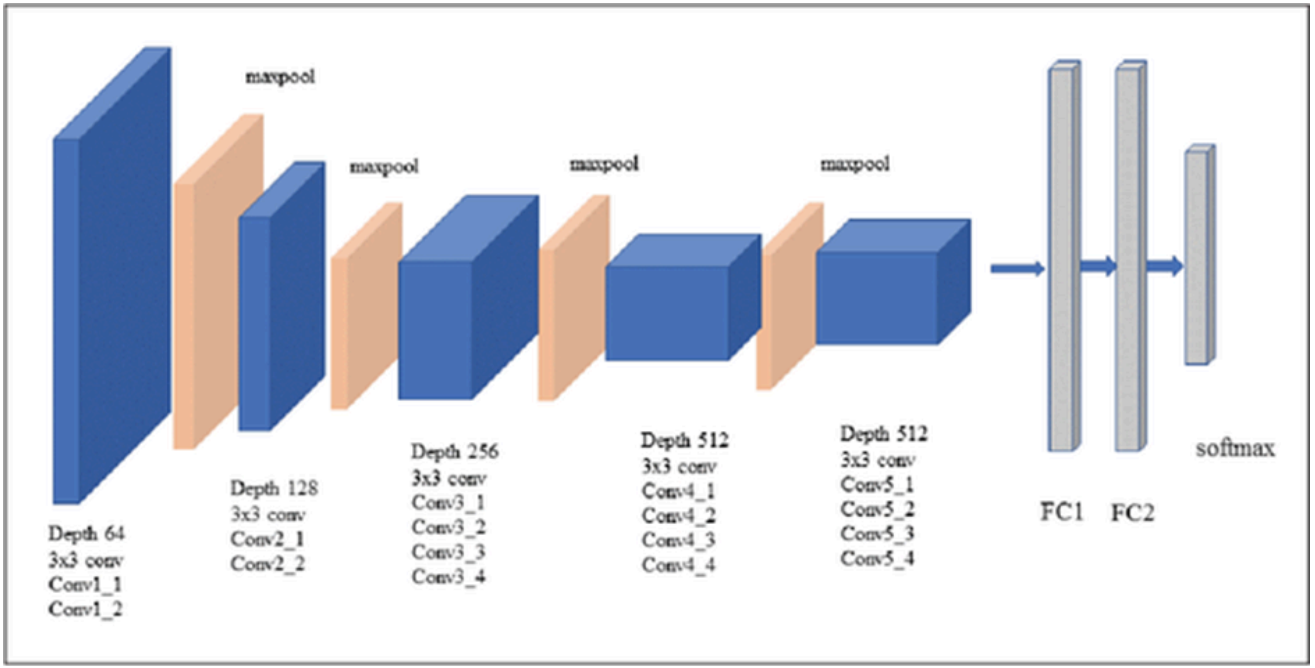

*Fig. 7: VGG network architecture*

## 5. Performance Results and Analysis

**Results of FGSM and CW Attacks and Defensive Distillation**

The objective of this research was to assess the impact of two distinct adversarial attack methods, namely FGSM and CW attack, on three widely used pre-trained CNN architectures: resnext50_32xd, Densenet201, and VGG19. These architectures, readily accessible in PyTorch's torchvision package, were initially trained on the ImageNet dataset. Additionally, the study explored the efficacy of Defensive Distillation in mitigating the effects of these attacks. Defensive Distillation, a method aimed at enhancing model robustness against adversarial perturbations, was examined as a potential defense strategy in this context. The investigation sought to provide insights into the vulnerabilities of these CNN models and the effectiveness of Defensive Distillation in bolstering their resilience to adversarial attacks.

**Classification Performance Before Attack**

We chose to use the Tiny ImageNet dataset, which has 200 classes, for our research because the ImageNet dataset has 1000 classes. We took this decision in order to better meet our computing limitations and research goals. Since it would not always be possible to assign a single, unique label to a picture from the Tiny ImageNet classes, we used both top 1 and top 5 accuracy scores. The percentage of successfully categorized images, when the top prediction made by the model matches the actual label, is indicated by the top 1 accuracy. To give a more complete picture of the model's predictive ability, we also evaluated the top 5 accuracy because of the wider range of possibilities found in the top 5 forecasts. If the true label occurs in any of the top 5 projected classes by the model, the image is deemed successfully identified according to this criterion. We attempted to convey the subtlety of image classification within the limitations of the Tiny ImageNet dataset by combining both top 1 and top 5 accuracy metrics.

*Table 1. The classification models' performance in the absence of an attack*

| Metric | Resnext50_32x4d Model | DenseNet201 Model | VGG-19 Model |
|---|---|---|---|
| Top-1 error | 10.16% | 13.92% | 19.88% |
| Top-5 error | 1.20% | 2.22% | 4.38% |

Three classification models were tested using the large Tiny ImageNet dataset, which has 200 different image classes. The performance results are shown in Table 1. The low top-% error rates of all three models indicate their remarkable precision. The Resnext50_32x4d model, in particular, has exceptional performance, combining the lowest error rates with the highest precision. Specifically, this model's Top-1 and Top-5 error rates are 10.16% and 1.20%, respectively.

After a thorough analysis of all three models on the Tiny ImageNet dataset, we looked at individual photos from the dataset. In order to achieve this, we investigated the classification results of the models using a random selection of photos indexed at 12, 18, and 23.

Interestingly, the classifications "great white shark," "tiger shark," and "hammerhead" are represented by these pictures, in that order.

*Table 2. The selected photos' categorization outcomes using the ResNext50_32x4d model*

| Image Index | Image True Class | Top-5 Predicted Classes | Predicted Top-5 Confidences |
|---|---|---|---|
| 12 | great white shark | great white shark<br>hammerhead<br>tiger shark<br>killer whale<br>submarine | 0.8236<br>0.0924<br>0.0824<br>0.0002<br>0.0001 |
| 18 | hammerhead | hammerhead<br>tiger shark<br>great white shark<br>gar<br>barracouta | 0.9935<br>0.0032<br>0.0021<br>0.0002<br>0.0002 |
| 23 | tiger shark | tiger shark<br>gar<br>eel<br>hammerhead<br>sturgeon | 0.9677<br>0.0092<br>0.0041<br>0.0031<br>0.0029 |

The results of the resnext50_32x4d model's classification of these three given photos are shown in Table 2. With the exception of the image of the "great white shark," all cases show very excellent categorization accuracy, with confidence levels for the real class reaching 90%. In this case, "confidence" refers to the likelihood that the model attributes to the relevant class. For example, the resnext50_32x4d model assigns a confidence rating of 0.8236 when it correctly predicts the class "great white shark" for an image. This indicates that there is a high probability (82.36%) that the image belongs to the "great white shark" class.

The classification results for the three photos using the resnext50_32x4d model are displayed in Fig. 8. The input image is shown on the left of the figure, and the model's confidence values for the top five predicted classes are shown on the right. Horizontal bars represent these confidence values.

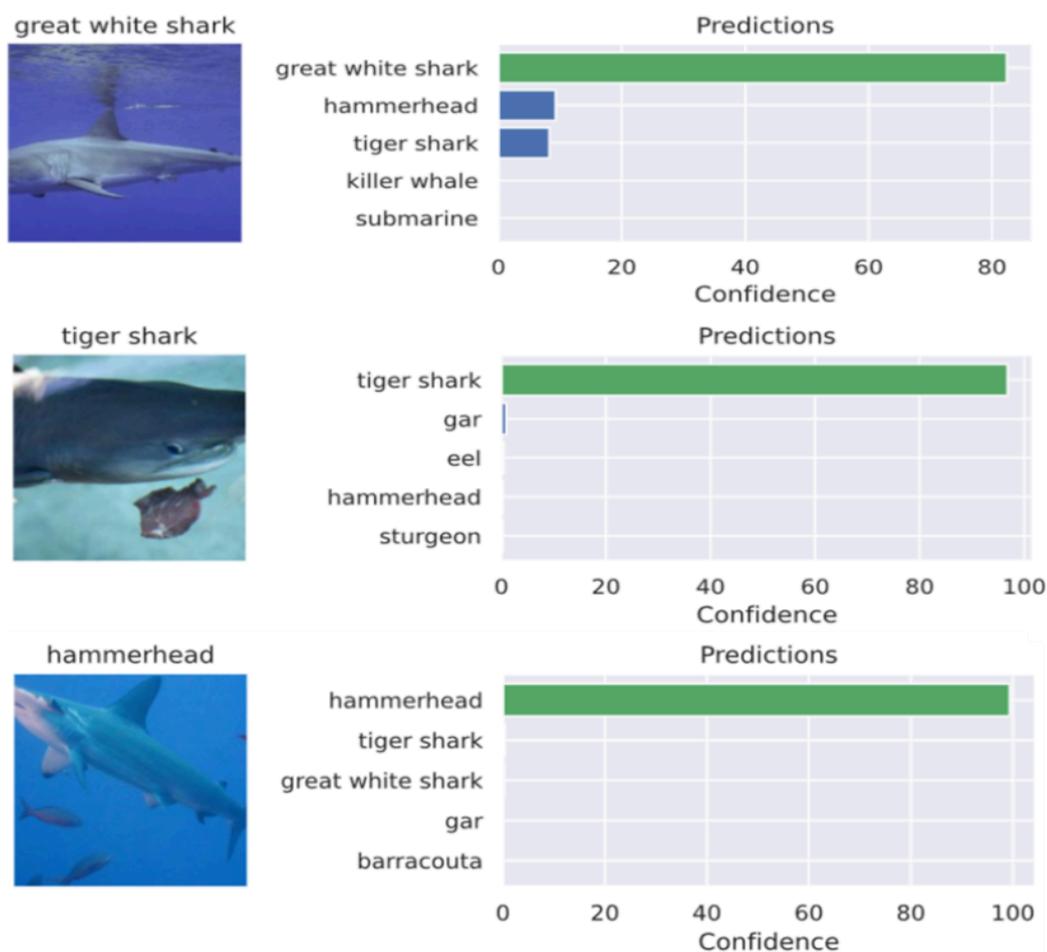

*Fig. 8: A few particular photographs' categorizations using the Resnext50_32x4d model*

**Classification Under The Influence Of The FGSM Attack**

Initially, we employed the FGSM attack with an epsilon value of 2%, indicating that pixel values were perturbed by approximately 1 within the range of 0 to 255. The perturbation was deliberately subtle, rendering the altered image indistinguishable from the original. Our focus was on evaluating the performance of the resnext50_32x4d model under this attack scenario with ε = 2% in Table 3.

Even with a low ε value of 0.02, the FGSM attack significantly compromises the performance of the resnext50_32x4d model. Furthermore, discerning between the adversarial images and the original ones proves to be a challenging task. The visual representation of the attack on the images is shown in Fig 9.

*Table 3. The selected images' classification outcomes for the resnext50_32x4d model under FGSM attack with epsilon = 2%*

| Image Index | Image True Class | Top-5 Predicted Classes and Confidences | |
|---|---|---|---|
| | | Class | Confidence |
| 12 | great white shark | hammerhead | 0.9886 |
| | | tiger shark | 0.0066 |
| | | great white shark | 0.0046 |
| | | gar | 0.0005 |
| | | sturgeon | 0.0001 |
| 18 | tiger shark | gar | 0.2853 |
| | | eel | 0.1596 |
| | | barracouta | 0.0639 |
| | | sea snake | 0.0591 |
| | | sturgeon | 0.0258 |
| 23 | hammerhead | tiger shark | 0.4591 |
| | | great white shark | 0.0686 |
| | | barracouta | 0.0670 |
| | | gar | 0.0400 |
| | | sturgeon | 0.0374 |

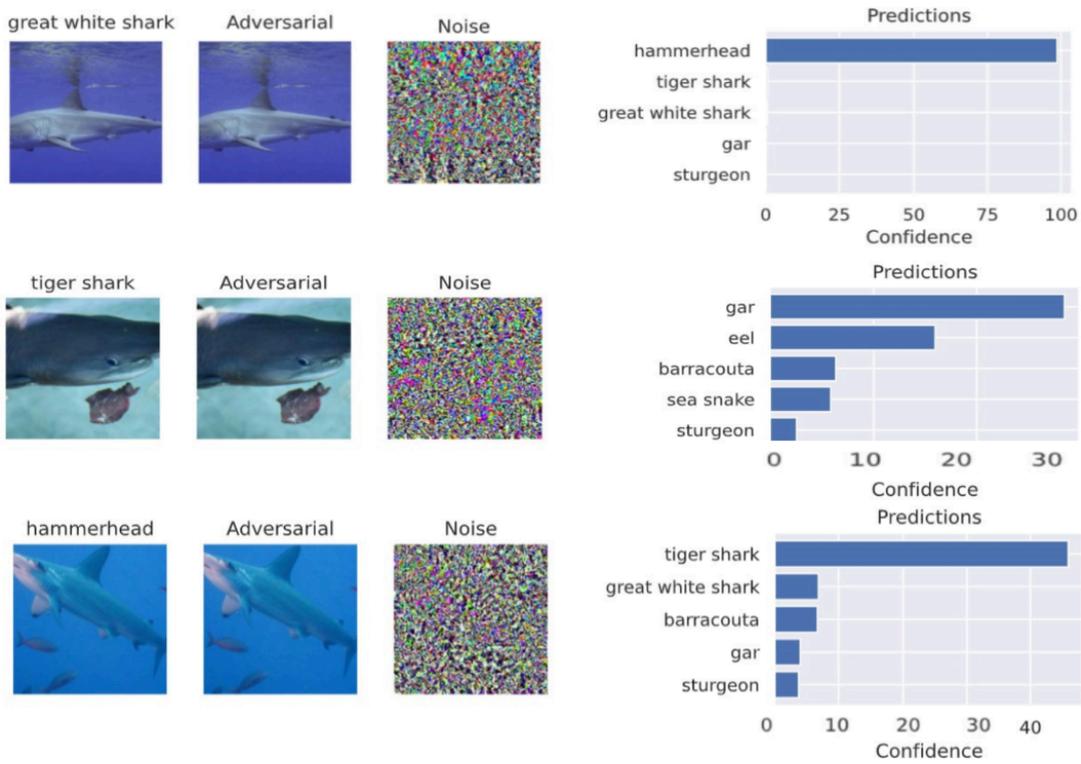

*Fig. 9: The resnext50_32x4d model's classification results under the FGSM attack with ε = 2%*

*Table 4. Performance of the ResNext50_32x4d Classification model under FGSM Attack for various values of ε*

| Noise Level (ε in %) | Top-1 Error (%) | Top-5 Error (%) |
|---|---|---|
| 1 | 77.88% | 33.82% |
| 2 | 87.62% | 49.58% |
| 3 | 90.34% | 55.62% |
| 4 | 91.38% | 59.14% |
| 5 | 91.80% | 60.58% |
| 6 | 91.64% | 61.36% |
| 7 | 91.34% | 61.66% |
| 8 | 91.16% | 61.60% |
| 9 | 90.96% | 61.58% |
| 10 | 90.74% | 61.16% |

*Table 5. DenseNet201 Classification Performance for Various Values of ε under FGSM Attack*

| Noise Level (ε in %) | Top-1 Error (%) | Top-5 Error (%) |
|---|---|---|
| 1 | 78.94% | 34.62% |
| 2 | 89.92% | 52.28% |
| 3 | 93.08% | 59.90% |
| 4 | 94.22% | 63.96% |
| 5 | 94.48% | 66.22% |
| 6 | 94.66% | 67.48% |
| 7 | 94.64% | 67.74% |
| 8 | 94.36% | 67.82% |
| 9 | 94.34% | 67.94% |
| 10 | 94.12% | 67.86% |

*Table 6. Performance of VGG19 Classification under FGSM Attack for Various Values of ε*

| Noise Level (ε in %) | Top-1 Error (%) | Top-5 Error (%) |
|---|---|---|
| 1 | 92.86% | 59.7% |
| 2 | 96.92% | 74.20% |
| 3 | 97.80% | 78.82% |
| 4 | 98.10% | 80.32% |
| 5 | 98.08% | 80.84% |
| 6 | 98.02% | 80.84% |
| 7 | 97.68% | 80.84% |
| 8 | 97.68% | 80.54% |
| 9 | 97.50% | 80.20% |
| 10 | 97.36% | 79.92% |

The epsilon value was incremented from 1% to 10% in steps of 1%. Throughout this progression, it was observed in Fig.10 that the error rate steadily increased from 0.01 to 0.04, after which it reached saturation. For the resnext50_32x4d model, the highest classification errors were recorded as Top-1 Error 91.80% and Top-5 Error 61.66%. Similarly, for the Densenet201 model, the errors were Top-1 Error 94.66% and Top-5 Error 67.94%. In the case of the VGG19 model, the errors peaked at Top-1 Error 98.10% and Top-5 Error 80.84%. The resnext50_32x4d, DenseNet201, and VGG19 models underwent a total of 157 iterations each, with average iteration times of 2.35 seconds, 2.21 seconds, and 1.65 seconds, respectively. The error values after the FGSM attack in percentage for different ε and the mentioned three models are presented in Table 4, Table 5, and Table 6.

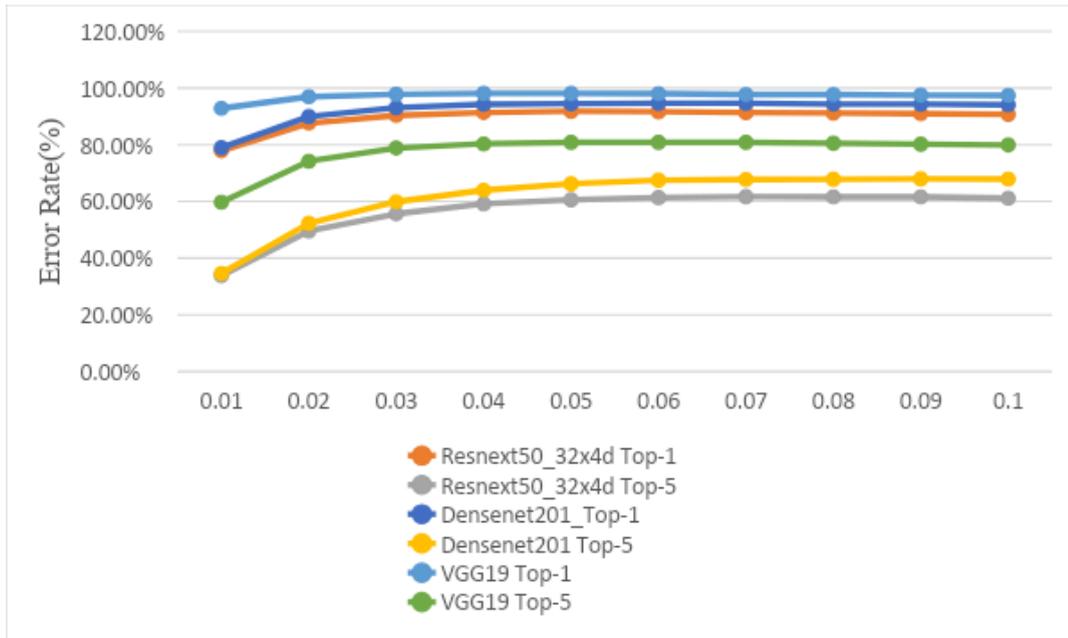

*Fig.10: Following an FGSM attack, the top-1 and top-5 error rates (%) for each model for a range of epsilon values*

**Classification Performance Under The CW Attack**

The CW attack was implemented with an epsilon value of 2%, perturbing pixel values by approximately 1 within the range of 0 to 255. This subtle perturbation was deliberately designed to make the altered image visually indistinguishable from the original. Our primary focus was to assess the performance of the resnext50_32x4d Model under this attack scenario, specifically with ε = 2%, as presented in Table 7.

*Table 7. Classification Performance of ResNext50_32x4d under CW Attack*

| Image Index | Image True Class | Top-5 Predicted Classes and Confidences | |
|---|---|---|---|
| | | Class | Confidence |
| 12 | great white shark | hammerhead | 0.9886 |
| | | tiger shark | 0.0066 |
| | | great white shark | 0.0046 |
| | | gar | 0.0005 |
| | | sturgeon | 0.0001 |
| 18 | tiger shark | gar | 0.2853 |
| | | eel | 0.1596 |
| | | barracouta | 0.0639 |
| | | sea snake | 0.0591 |
| | | sturgeon | 0.0258 |
| 23 | hammerhead | tiger shark | 0.4591 |
| | | great white shark | 0.0686 |
| | | barracouta | 0.0670 |
| | | gar | 0.0400 |
| | | sturgeon | 0.0374 |

Even with the relatively low ε value of 0.02, the impact of the CW attack on the resnext50_32x4d model's performance was substantial. The model's classification accuracy was significantly compromised under this attack. Moreover, distinguishing between the adversarial images and the original ones posed a considerable challenge. Despite the seemingly minor perturbations, the CW attack effectively undermined the model's robustness and highlighted the vulnerability of the resnext50_32x4d model to adversarial attacks. Furthermore, the results are graphically presented in Fig 11.

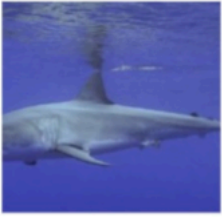
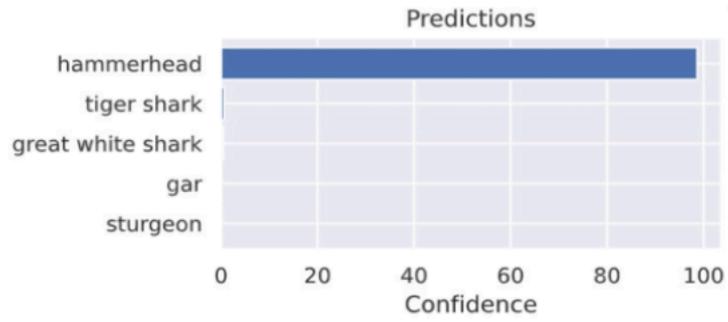
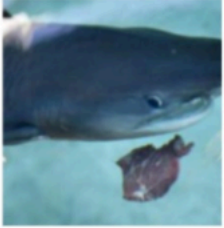
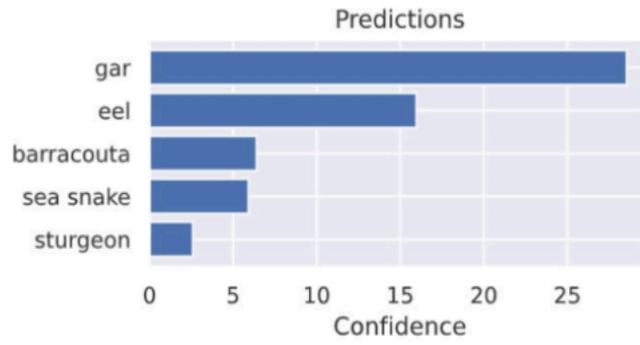
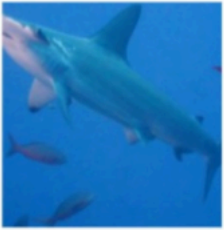
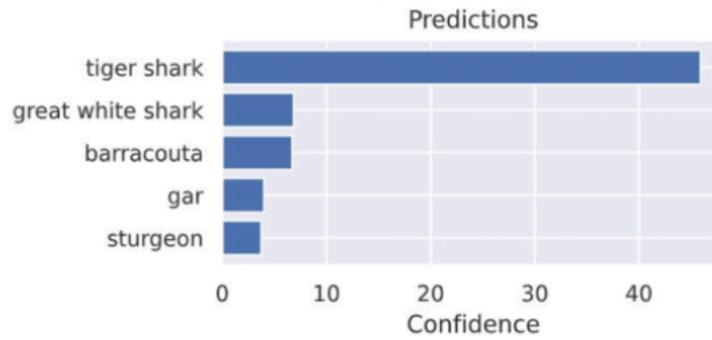

*Fig. 11: The classification results under CW assault with epsilon = 2% of a subset of particular images of the ResNext50_32x4d model*

*Table 8. ResNext50_32x4d Classification Performance for Various Values of ε under CW Attack*

| Noise Level (ε in %) | Top-1 Error (%) | Top-5 Error (%) |
|---|---|---|
| 1 | 77.88% | 33.86% |
| 2 | 87.62% | 49.58% |
| 3 | 90.34% | 55.62% |
| 4 | 91.38% | 59.14% |
| 5 | 91.80% | 60.58% |
| 6 | 91.64% | 61.38% |
| 7 | 91.34% | 61.66% |
| 8 | 91.16% | 61.60% |
| 9 | 90.96% | 61.58% |
| 10 | 90.74% | 61.16% |

*Table 9. Classification Performance of DenseNet201 under CW Attack for Various Values of ε*

| Noise Level (ε in %) | Top-1 Error (%) | Top-5 Error (%) |
|---|---|---|
| 1 | 78.94% | 34.64% |
| 2 | 89.92% | 52.28% |
| 3 | 93.08% | 59.90% |
| 4 | 94.22% | 63.96% |
| 5 | 94.48% | 66.22% |
| 6 | 94.66% | 67.48% |
| 7 | 94.64% | 67.74% |
| 8 | 94.36% | 67.82% |
| 9 | 94.34% | 67.94% |
| 10 | 94.12% | 67.86% |

*Table 10. VGG19 Classification Performance for Various Values of ε under CW Attack*

| Noise Level (ε in %) | Top-1 Error (%) | Top-5 Error (%) |
|---|---|---|
| 1 | 92.86% | 59.76% |
| 2 | 96.92% | 74.20% |
| 3 | 97.80% | 78.82% |
| 4 | 98.10% | 80.32% |
| 5 | 98.08% | 80.84% |
| 6 | 98.02% | 80.84% |
| 7 | 97.68% | 80.84% |
| 8 | 97.68% | 80.54% |
| 9 | 97.50% | 80.20% |
| 10 | 97.36% | 79.92% |

As a result, it can be seen that all three of the models in Figure 12 perform terribly when it comes to classification when subjected to the CW attack. For the resnext50_32x4d model, the highest classification errors were recorded as Top-1 Error 91.80% and Top-5 Error 61.66%. Similarly, for the Densenet201 model, the errors were Top-1 Error 94.66% and Top-5 Error 67.94%. In the case of the VGG19 model, the errors peaked at Top-1 Error 98.10% and Top-5 Error 80.84%. The resnext50_32x4d, DenseNet201, and VGG19 models underwent a total of 157 iterations each, with average iteration times of 24.9 seconds, 27.4 seconds, and 34.7 seconds, respectively. The error values after the CW attack in percentage for different ε and the mentioned three models are presented in Table 8, Table 9, and Table 10.

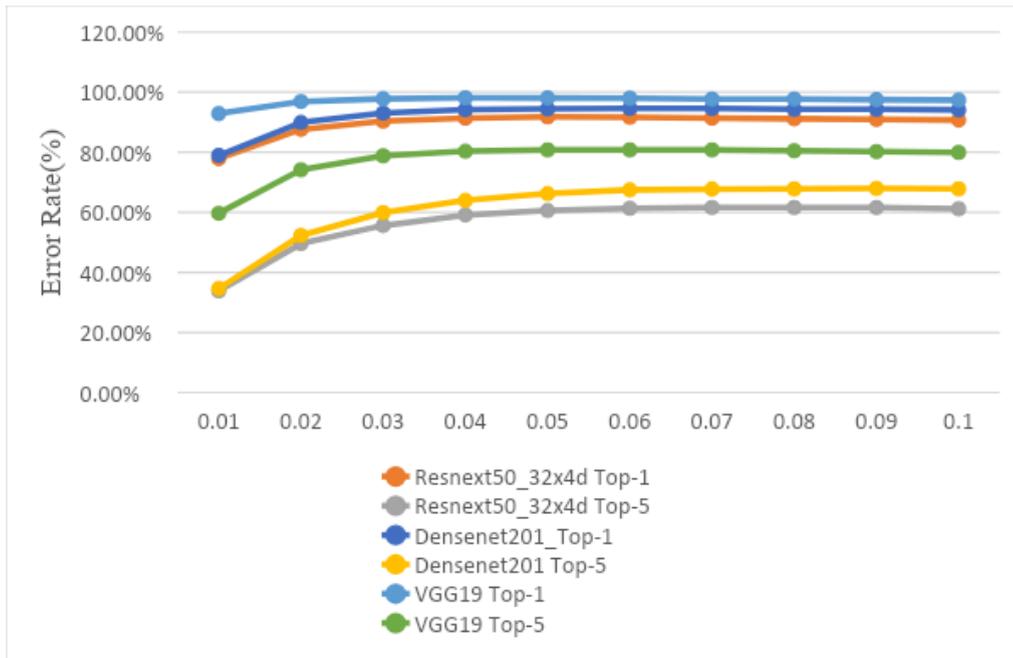

*Fig.12: Top-1 and Top-5 error rates (%) for various epsilon values following CW assault in all models*

**Performance of Defensive Distillation on FGSM Attack**

The CIFAR-10 dataset comprises 60,000 32x32 color images categorized into 10 classes, each containing 6,000 photos. This dataset served as the basis for investigating the impacts of the FGSM attack and assessing the defensive distillation technique's efficacy in mitigating the attack. Following this, a CNN model was trained for image classification using a randomly selected subset of 40,000 samples, with an additional 10,000 samples reserved for validation. Subsequently, the model's performance was evaluated using the remaining 10,000 samples from the dataset.

Distillation was originally used to reduce a huge model (called the instructor) to a smaller version (called the distilled model). Training the distilled model with these soft labels rather than hard labels taken straight from the training set entails first training the teacher model on the dataset, and then applying soft labels to instances based on the teacher's output vector. This method increases the efficiency with which the distilled model learns to predict challenging labels and improves the accuracy of the test dataset. The instructor model in this case is resnet101, while the student, or distilled, model is resnext50_32x4d.

Using the Adam optimizer, an adaptive optimization approach, the parameter temperature is adjusted to 100 and the instructor model is trained on the CIFAR-10 dataset. The optimizer uses beta parameters (0.9, 0.99) to control the exponential moving means of gradients and squared gradients, and it uses a minimum learning rate of 0.0001. Because they accommodate different gradients across parameters, these parameters allow the neural network to dynamically modify the learning rates for each parameter, resulting in smoother convergence during training.

It uses the categorical cross-entropy loss function. The validation loss is tracked by a learning rate scheduler, which lowers the learning rate when the tracked parameter reaches saturation. The *mode* parameter of the scheduler is set to a *minimum* which results in a drop in the learning rate when the validation loss stops declining. Since the *factor* parameter is equals to 0.1, when the validation loss reaches a saturation, the learning rate will be lowered by a factor of 0.1. Additionally, the parameter *patience* is set to 3, which indicates how many epochs to wait in case the monitored measure stalls before modifying the learning rate. If, throughout the allotted patience period, the monitored metric does not show improvement, the scheduler will reduce the learning rate; in this case, after three epochs. Ten training epochs are applied to the model. The teacher model's training and validation losses are depicted in Fig. 13(a), while the student model is plotted against the number of epochs in Fig. 13(b).

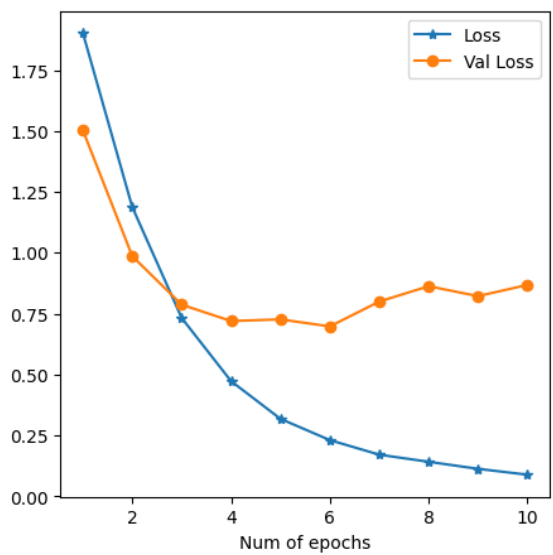
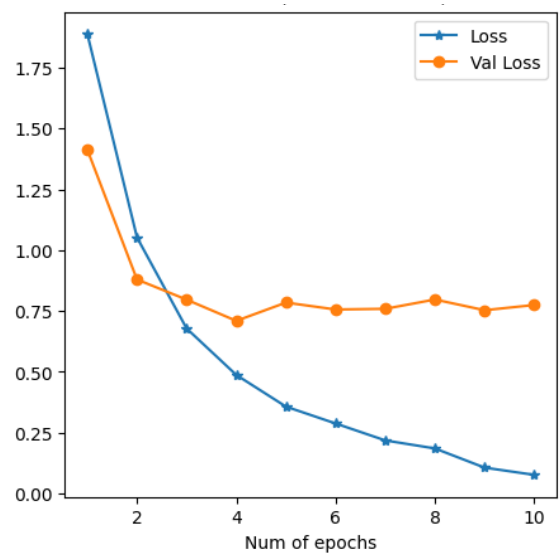

Fig. 13(a): The instructor model's training and validation loss trends over epochs are shown

Fig. 13(b): The student model's training and validation loss trends throughout epochs are displayed

The defensive distillation models underwent an FGSM attack using perturbed adversarial samples across various epsilon values (epsilon = [0%, 0.7%, 1%, 2%, 3%, 5%, 10%, 20%, 30%]). The classification accuracy before and after the attack, with and without defensive distillation, is depicted in Fig.14. The accuracy before defensive distillation is shown by the green line, while the accuracy post-defensive distillation is indicated by the blue line. Initially, the accuracy of the resnext50_32x4d model, utilized as the student model in defensive distillation and trained on the CIFAR10 dataset, was 0.79. After the attack, the accuracy dropped to 0.55, but following distillation, it improved significantly to 0.87.

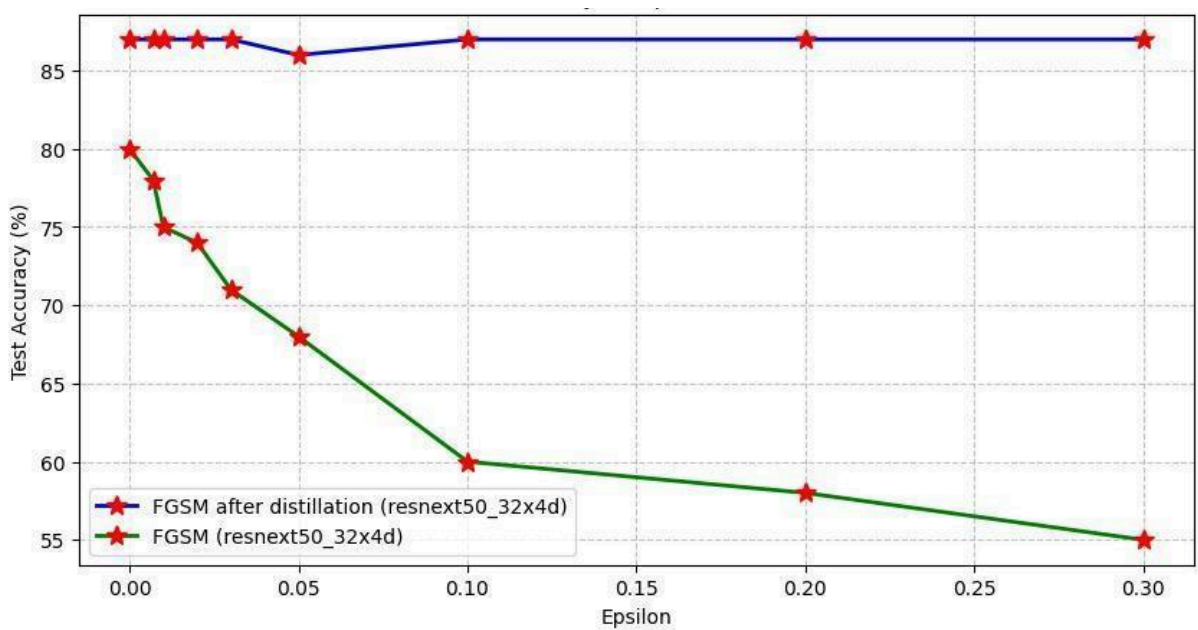

*Fig. 14: The ResNext50_32x4d model's accuracy values throughout a range of epsilon values, both before and after defensive distillation*

**Performance of Defensive Distillation on CW attack**

The defensive distillation models were subjected to a CW attack utilizing perturbed adversarial samples across a range of epsilon values ([0%, 0.7%, 1%, 2%, 3%, 5%, 10%, 20%, 30%]). Fig.15 illustrates the classification accuracy both before and after the attack, with and without employing defensive distillation. The accuracy prior to applying defensive distillation is represented by the green line, while the accuracy following defensive distillation is denoted by the blue line. The graph indicates that defensive distillation did not improve accuracy after the CW attack. Notably, both teacher and student models were trained on the CIFAR-10 dataset, with parameters set under the same conditions as the FGSM attack.

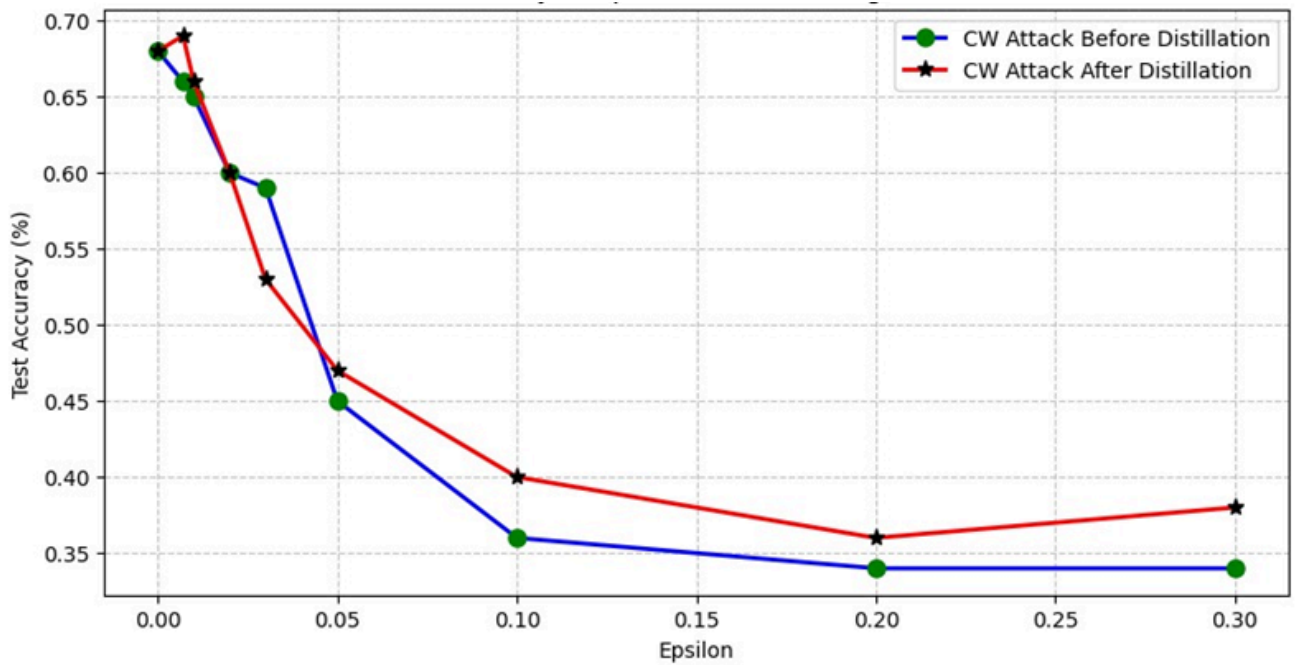

*Fig. 15: The ResNext50_32x4d (student) model's accuracy values before and after defensive distillation for a range of epsilon values*

# 6. Conclusion and Future Work

Even if our defensive distillation model successfully thwarts attacks like FGSM, it is still vulnerable to more sophisticated techniques like the CW attack. However, our findings highlight the possible effectiveness of defensive distillation against adversarial tactics such as FGSM. In the future, it will be crucial to improve the model's defensive capabilities by adding a richer, more diverse dataset to it. By strengthening the model's robustness, this tactical improvement seeks to increase its usefulness in defending against adversarial attacks in picture classification tasks.

Protecting image classifiers from adversarial attacks is an important area of research in artificial intelligence, with concerns about the effectiveness of defense strategies such as distillation approaches. Defensive distillation has been implemented to improve robustness against attacks; nonetheless, it has been demonstrated that this approach is insufficient against well-known adversarial tactics like the CW attack. This emphasizes how urgently new defense mechanisms are needed to effectively counter the dynamic danger landscape in artificial intelligence.

We have also compared the results of the CW L2 attack and FGSM attack and have seen the iterative taken for the former is more than that of the latter. Also, the test errors (top-1 and top-5) are mostly the same in both cases with an increase in perturbation. This provides us with a scope of working deeply into why it happens in the future.

Researchers need to use strategies other than traditional defenses, such as adversarial training and robust optimization, to counter adversarial attacks on image classifiers. It is imperative to fortify current techniques, such as defensive distillation, against a range of assault strategies, including FGSM. To strengthen image classifiers' resilience in the face of the constantly changing AI adversarial threat scenario, this collaborative effort is essential.

To sum up, the competition between defensive measures and adversarial attacks highlights the intricacy and dynamic nature of the subject of adversarial machine learning. Deep learning model reliability is seriously threatened by assaults like FGSM and CW L2, although methods like defensive distillation present viable ways to increase model robustness. It is crucial to explore multidisciplinary methods that integrate knowledge from machine learning, optimization, and cognitive science as this field of study develops further to create AI systems that are more reliable and durable. In the end, combating adversarial attacks necessitates a thorough comprehension of the fundamental weaknesses of neural networks as well as the creation of all-encompassing defense plans that give equal weight to security and accuracy.